# Computational Intelligence for Qualitative Coaching Diagnostics: Automated Assessment of Tennis Swings to Improve Performance and Safety

Authors:

Boris Bačić[a,b] and Patria A. Hume[b]

[a]*School of Engineering, Computer and Mathematical Sciences, Auckland University of Technology, Auckland, New Zealand;*
[b]*Sport Performance Research Institute New Zealand, Auckland University of Technology, Auckland, New Zealand*

**Corresponding author:** Dr Boris Bačić, boris.bacic@aut.ac.nz

**Abstract**

Coaching technology, wearables and exergames can provide quantitative feedback based on measured activity, but there is little evidence of qualitative feedback to aid technique improvement. To achieve personalised qualitative feedback, we demonstrated a proof-of-concept prototype combining kinesiology and computational intelligence that could help improving tennis swing technique utilising three-dimensional tennis motion data acquired from multi-camera video. Expert data labelling relied on virtual 3D stick figure replay. Diverse assessment criteria for novice to intermediate skill levels and configurable coaching scenarios matched with a variety of tennis swings (22 backhands and 21 forehands), included good technique and common errors. A set of selected coaching rules was transferred to adaptive assessment modules able to learn from data, evolve their internal structures and produce autonomous personalised feedback including verbal cues over virtual camera 3D replay and an end-of-session progress report. The prototype demonstrated autonomous assessment on future data based on learning from prior examples, aligned with skill level, flexible coaching scenarios and coaching rules. The generated intuitive diagnostic feedback consisted of elements of safety and performance for tennis swing technique, where each swing sample was compared with the expert. For safety aspects of the relative swing width, the prototype showed improved assessment (from 81% to 91%) when taking into account occluded parts of the pelvis. The next-generation of augmented coaching, personalised rehabilitation and exergaming systems, based on the concepts presented, will be able to help improve end-user's sport discipline-specific techniques by learning from small expert-labelled data sets, adapting and providing personalised intuitive autonomous assessment and diagnostic feedback aligned with a specified coaching programme and context requirements.

**Introduction**

The emerging area of wearable and ubiquitous computing devices is expanding into many applications, notably in domains of physical activity, sport, rehabilitation, exergames and healthcare such as Babolat (http://www.babolatplay.com), HOCOMA (http://knowledge.hocoma.com/research.html) and Zepp (http://www.zepp.com). Increasingly popular, such *augmented coaching systems and technology* (ACST) is becoming accessible to global consumer population and can produce, process and exchange large amount of motion data before providing feedback based on measured human activity (Boris Bačić, 2016c; Lightman, 2016). At present, it is common knowledge that ACST and exergames are not able to advise users how to improve their motor learning, swing technique or general motion technique in particular sport or rehabilitation domains (Boris Bačić, 2016c). To illustrate ACST operation exchanging and processing demands on increasingly larger motion data that are typically exchanged on-line and by using a cloud service provider, we may consider a hypothetical example of a recreational runner, who is using a small action camera, a mobile phone, smart watch and wearable sensors. In addition to keeping a record of past activities, such technology can fuse data from sensors and video to collectively augment video replay by overlaying information with telemetry data (speed, map location), pace, time lapse, number of steps (from start, per minute), heart rate, heel, toe or mid-shoe strike and average pronation angle (jogging, sprinting), comparisons with selected running communities (running speed and efficiency, shock absorption rate), running, walking and resting time. Although current technology is able to process raw data from sensors and video and provide visualisation and other reporting capabilities on measured quantitative information unfortunately, it is not able to help end-users to improve their running technique, or to help them to adapt their personal running style and footwork technique for soccer or tennis. However, by combining the quantitative diagnostic information produced, with expert's help (e.g. athletic or soccer coach), an end-user may still be able to improve his/her running technique following recommendations from a coach - such as *'hold your hips high!'* or *'lift your knees'* and conclude the training session with *'pronating is good but over pronation isn't'*. This coach-to-athlete feedback is known as verbal cues, which as a communication is: (1) immediate, short and informed by collected data, (2) personalised and directed to drive attention focus of the athlete, while other bystanders might not be familiar with the meaning of such attention cues and (3) such cues may also not be applicable to other athletes. What a coach can also recommend from diagnosis based on collected quantitative data, prior knowledge about the athlete, assessment criteria and observed video is a set of intervention exercises aimed to help the athlete to achieve his/her personalised goals (D. V. Knudson, 2013).

Although the idea and notion of applying AI to analysis of sport performance is not new (B. Bacic, 2006; Lapham & Bartlett, 1995), current ACST can assess *knowledge of results* (KR) and provide quantitative information about the movement. In general, ACST is expected to promote physical activities and in some cases, biomechanical analysis above human cognitive abilities. To enable advancements towards the next generation of ACST that can help to improve sport-specific technique such as complex and personalised tennis swings, this study addressed the following

questions:

(1) Is it possible to design a system that is capable to assess human performance, translate derived quantitative information from raw motion data and provide personalised qualitative feedback similar to a coach, even if information of the movement outcome or KR is not available?

(2) Given that expert labelling on a large dataset is not feasible, is it possible to design a system that can learn from small training data?

Existing qualitative models (Dunham, 1994; Gangstead & Beveridge, 1984; Hay & Reid, 1984) share subjective/descriptive rules related to *knowledge of performance* (KP) that guide assessment of observed motion and general notion of temporal and spatial analysis. In agreement with Knudson and Morrison (2002, p. 132), when the expert familiar with common errors is confident that "a critical feature related to body motion needs improvements, the research suggests that KP feedback is more effective". Some of the rules and critical features of human movement can be quantified and communicated in biomechanical terms such as knee flexion angle with categories representing optimal and sub-optimal ranges that are common for a specific motion pattern (D. V. Knudson & Morrison, 2002, p. 81). However, for complex motion patterns it is also known that abstract and rather descriptive common-sense rules may not be easy to define, identify, validate and assess numerically as the optimal ranges and their *if-then-else* rules or as fuzzy rules (Kecman, 2001; Zadeh, 1965).

Regarding abstract coaching rules and technique improvements using the *attention cuing* method during drill practise, a coach may say a few cue words to a player who has previously learned their meaning. In open-skill sports such is tennis, there are individual variations of complex movement patterns (Bollettieri, 2001; Hughes & Bartlett, 2002; Reid, Elliott, & Crespo, 2013). Given the winning objective, players often need to demonstrate versatility and adaptability to multiple goals of human movement (e.g. performance in imparting energy, efficiency, efficacy, consistency, movement robustness, recovery time for next response, balance, and safety) that are all directed at neutralising and responding to opponent's actions, environment and other circumstantial constraints. For many sport disciplines, such as tennis, the game concepts have evolved in a way that the 'old-school wooden racquets' coaching is not applicable to e.g. faster balls, shorter preparation times and on-going technological advancement of sport equipment (Bačić, 2018). Given the need for flexibility of assessment criteria of multiple goals of human movement that may also be mutually conflicting and subject to skill-level expectation, the next-generation of ACST should mimic and synthesise some of the coaches' abilities to prioritise and personalise feedback. For the design and development of such ASCS systems, anticipated capabilities should include capturing initial coaches' assessment decisions, learning from initially small decision and data sets, adapting and evolving its internal operation for future data even if motion data is captured beyond human cognitive abilities (e.g. in high frame rates or in sub-millimetre precisions).

For sport and rehabilitation science, expressing motion patterns from human movement into the *n*-dimensional mathematical spaces allows development of classification or prediction models that can establish fit of observed motion patterns into descriptive categories or ranges similar to a coach who provides KP feedback to aid technique improvements.

The challenge from increasingly growing human motion datasets, lack of personalised immediately available expertise (similar to a coach, medical/rehabilitation specialist), and the technology gap that artificial intelligence (AI) should address, is to emulate the human cognitive ability to translate and assess the data collected into actionable and meaningful advice for end-users that can be visualised and systematically organised into problem areas that are dependent on data availability and complexity (Fig. 1a). To illustrate design complexity involved in developing a system that can provide meaningful feedback using a tennis case study, such system would need to be incrementally developed from prior work. The prior work on tennis activity pattern recognition and diagnostic classification capabilities include: (1) swing recognition ('forehand' and 'backhand'); (2) swing phasing analysis ('preparation', 'action zone' and 'follow through'); (3) categorising ball hitting stance angle relative to intended target line ('closed', 'squared', 'semi-open' and 'open' stance); and (3) flexible skill-level assessments of 'good' or 'bad' swings. For current sensor-based systems 'good' and 'bad' swing classification could possibly be implemented in a simple fashion e.g. based on impact vibration pattern correlated to ball impact at the racquet's sweet spot – however for exergame development there is no ball interaction and sometimes swings executed with good technique may still missed the sweet spot (or vice versa, where 'bad' swings can still hit the sweet spot). Furthermore, 'good' or 'bad' swing classification may also not be easily implemented relying on traditional algorithmic approaches such simple descriptions of biomechanical parameters such as anatomical joint angles movements in time, top-spin and impact velocity but should rather be inspired by human cognitive ability to develop 'feel' for the racquet movement, which can be implemented into an AI-based system.

The underlying technological foundation for this study (Fig. 1 b), includes experimental developments of *design science* artefacts (Kampenes, Anda, & Dyba, 2008; Mingers, 2001; Runeson & Höst, 2009). The produced artefacts include architecture, temporal and spatial automated analysis of forehands, a sliding-window approach with kinematic parameters processing (B. Bacic, 2004; Boris Bačić, 2016b; Boris Bacic, Kasabov, MacDonell, & Pang, 2008); classifications od groundstrokes and stances (Boris Bačić, 2016b), the use of traditional and evolving *artificial neural networks* (ANN) (B. Bacic, 2006; Boris Bačić, 2013); 'good' and 'bad' swing classification based on computed racquet's sweet spot, *vector flow* feature extraction technique to emulate 'feel' and combined with a produced ANN-based system (Bačić, 2018). More complex solution similar to specialised brain regions working together inspired the recent application of the *reservoir computing* of ANN ensembles for swing detection, phasing analysis and identifying the intended ideal impact zone (> 90% accuracy) consisting of 2-3 frames captured at 50 Hz (Boris Bačić, 2016a, 2016d). Identifying the ideal impact zone of a swing is important for exergaming and rehabilitation contexts, where there is no statistical ground truth about ball impact, since the ball information is not recorded.

(a)

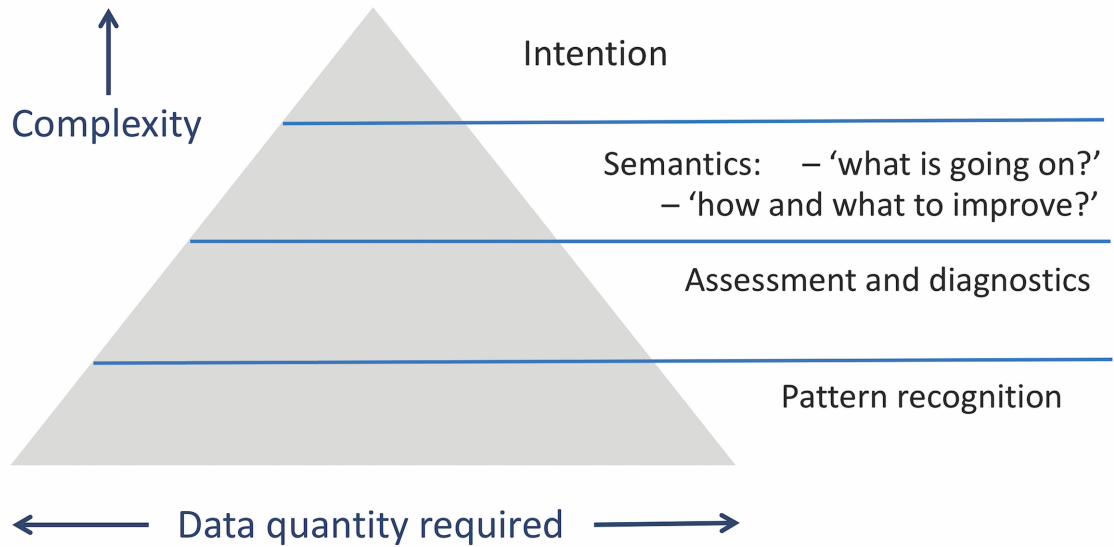

(b)

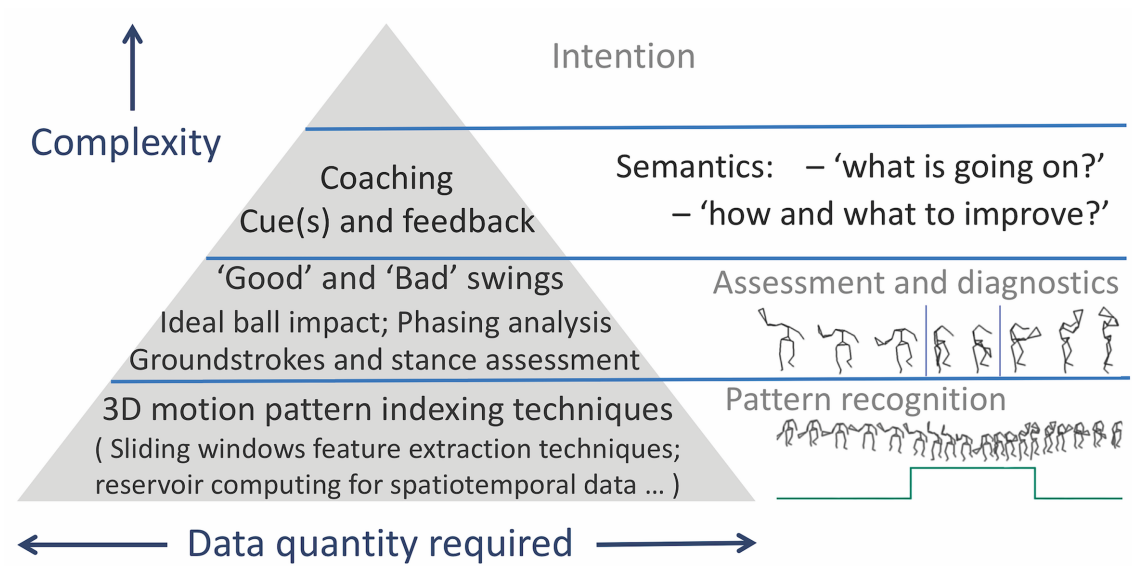

Fig. 1. Big data and AI challenges for: (a) human motion modelling and analysis for sport performance and rehabilitation; and (b) incremental development of artefacts leading to a personalised tennis coach prototype that can generate human-intelligible and common sense instruction for tennis technique improvements.

Based on prior work, the purpose of this multi-disciplinary study (Fig. 1b) was to demonstrate that it is possible to automate aspects of qualitative analysis of human motion including: (1) capturing expert's subjective insights that govern assessment criteria of (2) complex movement patterns, and (3) providing simple intelligible, intuitive personalised diagnostic feedback could aid technique and motor learning. The prototype system should be flexible and adaptive to take into account personalised idiosyncrasies such as skill-level and coaching scenarios (e.g. coaching drills and tasks) that focus on specific technique aspects. The system should also exhibit flexibility in assessment aligned with multiple objectives of human motion such as safety and performance and provide feedback in natural language (e.g. as intentional verbal cueing by a coach) and via augmented replay. For system requirements where traditional algorithmic approaches and traditional neural networks may not be able to deliver the best solution, our experimental design relies on methods from computational intelligence (CI), that a branch of AI includes modelling and development of adaptive and evolving systems (Włodzisław Duch & Mańdziuk, 2007; W. Duch, Setiono, & Zurada, 2004; N. Kasabov, 2007). Problem areas where evolving systems are expected to perform well include: incremental learning requirements; availability of small data sets; and where new data may include evolved patterns. Unlike evolving systems, traditional ANNs may suffer from performance degradation and with new data or even suffer from catastrophic forgetting – a phenomenon where "the system would forget a significant amount of old knowledge while leaning from new data" (N. Kasabov, 2007, p. 8).

**Materials and Methods**

For the purpose of prototype development, human motion data were captured to allow model design, and to present the developed constituent models and the user with a diversity of 'good' to 'bad' tennis strokes. The captured data set covered a variety of forehands and (single-hand) backhands that are typical at skill-levels from beginners to advanced-intermediate, including common errors (D. V. Knudson & Morrison, 2002, p. 219). The data set also contained a balanced distribution of forehands and backhands performed at diverse swing speeds, and from different hitting stances (Boris Bačić, 2016b). After multiple trial sessions, 43 swings were recorded in one session by a certified tennis coach (the first author) under guidance of another certified international tennis coach. Recorded swing samples were validated and independently assessed (100% agreement on 'good' 'average' and 'bad' swings) by two other professional New Zealand tennis coaches using a 3D stick figure player allowing virtual vantage point and zoom with 360° interactive observing angles during the replay (Boris Bačić, 2013, 2015). The tennis motion data set was captured in an indoor laboratory setup, using nine fixed-location cameras at 50 Hz using a SMART-e 900 eMotion/BTS motion capture system and exported into ASCII text data format for model design, prototyping, and 3D motion visualisation. The captured data contained multi-time series of 3D coordinates of a set of 22 retro-reflective markers. As shown in Fig. 2, the markers were attached to a racquet, the shoes and anatomical landmarks of the human body.

| Closed stance (to a minor degree) | Square stance (SqS) | Semi-open stance (SoS) | Open stance |
|---|---|---|---|
| 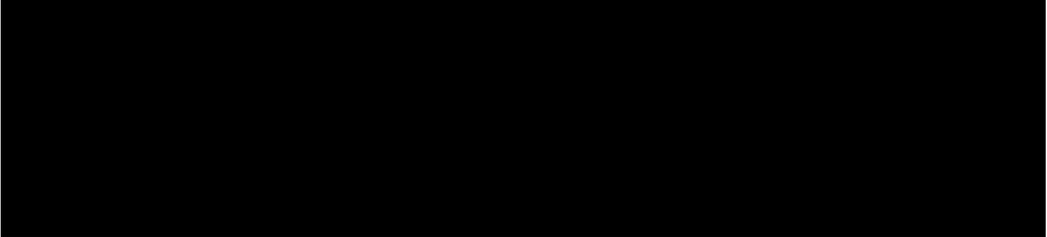 | | | |
| **SqS**: average<br>**SoS**: bad | **SqS**: very good<br>**SoS**: average | **SqS**: average<br>**SoS**: very good | **SqS**: bad<br>**SoS**: average |

Fig. 2. Visualisation of assessed stances. Two stance examples (SqS='Square Stance', and SoS='Semi-open Stance') show flexible coaching scenarios and subjective assessment categories ('very good', 'average', 'bad'). Adapted by permission from Springer Customer Service Centre GmbH: Springer Nature, Springer eBook "Extracting Player's Stance Information from 3D Motion Data: A Case Study in Tennis Groundstrokes", B. Bačić, copyright 2016.

The utilised minimalistic marker set layout (Fig. 2) was intended to have similar topology to Microsoft's Kinect stick figure model (MathWorks, 2016), but with extra markers to provide additional information (e.g. forearm pronation/supination for incremental research and development purposes).

For the objective of the study there were additional restrictions related to the captured data: (1) for the targeted skill-level, there were no open-stance single-hand backhands as encouraging single-hand backhand swings from open-stance was not considered a safe coaching practice (Boris Bačić, 2016b); (2) no synthetic, only human motion data were used in the study; and (3) regarding suppressing knowledge of results no ball or impact information were recorded.

For system and model developments, the chosen descriptive common-sense coaching rules associated with subjective and flexible assessment criteria were: stance, 'low-to-high' and swing width. From a tennis coaching perspective, the chosen coaching rules that are guiding subjective assessment criteria are also emphasised differently at each skill level or during drill practice in given coaching scenario. For example, a coach may prefer to introduce the 'square' (or 'side-on') stance as basic stance from the beginning, while discouraging other stances until a player progresses and improve their balance (Fig. 2). Some coaches prefer to introduce only three stance categories ('open', 'neutral' and 'closed') instead of four ('closed', 'square', 'semi-open' and 'open'). As the player progresses through the skill-levels, the semi-open and open stances are likely to be encouraged to improve response times. The stance selection may differ depending on additional factors such as ball bounce height and court coverage circumstances (Salzenstain, 2013). At the intermediate level, players may be expected to perform groundstrokes from all stances including optimal positions and positions while on the run, and hiding the intended target line. Regarding the 'low-to-high' rule, it is a cue used interchangeably with 'brush the ball', which a coach may typically say to a beginner to help improving their contact with the ball and to generate topspin (D. Knudson & Bahamonde, 2001). The coaching rule 'swing width' is typically not coached in isolation; given that players often extend their arm in the response situations such as during the full-reach return of serve. As the optimal contact zone depends on swing type, stance, grip, style, player's kinanthropometric profile and preferences, it is considered that 'swing width' assessment (as e.g. 'your swing is too wide') is also subjective. In addition, there is coaching rationale that a player should 'reach the ball with the feet', and 'swing width' feedback to be used occasionally only as a safety warning or with objectives to help player to improve swing control, consistency, response times or backswing motion (Bollettieri, 2001, pp. 54-56).

After identifying static and dynamic critical features of swing performance or safety for descriptive coaching rules, the next stages involved further data analysis, visualisation and biomechanical parameters transformations. Produced algorithms were able to transform critical features of swings as input data for classifier modelling and validation. Regarding the adaptive classifier model design, the chosen evolving classifier model was based on *Evolving Clustering Method* ECM (Nikola Kasabov, 2002, p. 41) and used as a neuro-fuzzy classifier known as *Evolving Clustering Function* ECF (Song et al., 2008). The benefit of the chosen ECF classification model was that it allowed incremental learning starting from the relatively small and unevenly balanced data set. In addition, it allowed generated machine knowledge extraction that could be used to initialise model operation as generated machine knowledge as a snapshot in time, therefore allowing adaptation to future extension of the original data set used for this study. Temporal and spatial problem analysis was expressed in finding frames in the *region of interest* (ROI) within the swing and transforming biomechanical values into a mathematical space that could be separated into distinct assessment categories by a classifier model. After motion data transformation and analysis, the produced critical features were used as machine learning features for classifier model developments. In general, for data sets containing variety of good, average and bad swing samples, obtaining relatively high OA would indicate that the adaptive system could utilise prior expert assessment data for automated assessment function that would also likely generalise well on future data. For model validation on the small data set, leave-one-out cross-validation method was considered as appropriate in order to avoid the possibility that a minority of data (of similar swing patterns) could be randomly selected into a training or a testing portion of the data set so in that case the produced system could not guarantee good generalisation performance on future data. Comparing to expert assessment, the *overall accuracy* (*OA*) was calculated (1) as:

$$OA = \frac{CS}{AS} \times 100 [\%] \qquad (1)$$

Where *CS* is the number of swing samples correctly categorised and *AS* is the number of all swing samples.

Regarding the expert assessment, all swing samples were visually assessed and categorised by the expert relying on 3D stick figure replays. In addition to the functional prerequisite of accurate 3D replay for expert validation and augmented feedback, the prototype – named as *Personalised Tennis Coaching System* (PTCS) – conveyed the following concepts: (1) training and validation of descriptive common-sense-based assessment modules; (2) flexible assessment criteria and (3) expert assessment and end-user feedback combining natural language qualitative feedback with smooth interactive and accurate replay of motion data. The user interface allowed choosing skill level, coaching scenario prioritised assessment and feedback visualisation. Minimalistic feedback in natural language for every swing was presented as a user-configurable list of weighted assessments to show verbal cues. The verbal cues as attention cues were colour-coded (for very good, average, and bad) and the list was sorted from worst towards the best critique (or praise).

**Results**

The results included the evidence of automated data transformation, analysis visualisation and processing integration in *graphical user interface* (GUI) to provide qualitative feedback of assessed tennis swings. For the PTCS prototype design, it was beneficial to combine two programming languages achieving additional GUI functionality and two-way communication between Delphi and the Matlab codebase. To improve interactive and graphic performance needed for validity and expert's swing assessment relying on the animated 3D stick figure replays, the original Matlab algorithms were manually transcoded in Delphi (Boris Bačić, 2015).

Regarding feature extraction for computer model development, it was not necessary to adhere strictly to three commonly used stages (preparation, action, follow-thorough) for temporal phasing and analysis (Gangstead & Beveridge, 1984). For relatively low sampling frequency (50 Hz), the motion data transformation algorithms were based on the coach's insight into what happens before and around the estimated action/impact zone. In case of 'low-to-high' and 'swing width' visualisation (Fig. 3), the region of interest (as temporal and spatial focus) was around the time when the hand with the racquet passed the pelvis region, influencing the rest of the swing movement.

(a) Tennis Swing in Sagittal Plane      (b) Zoom-in Region of Interest

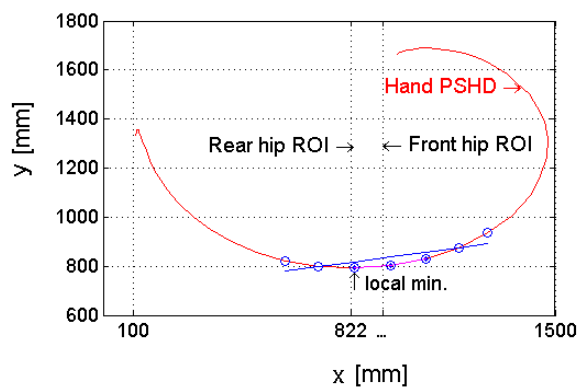
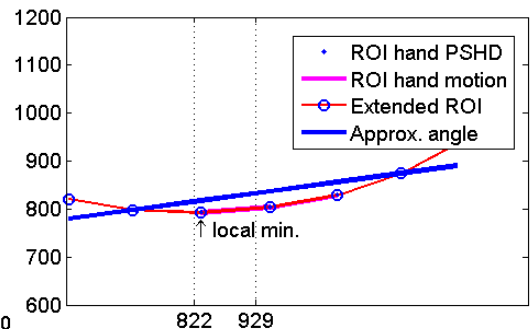

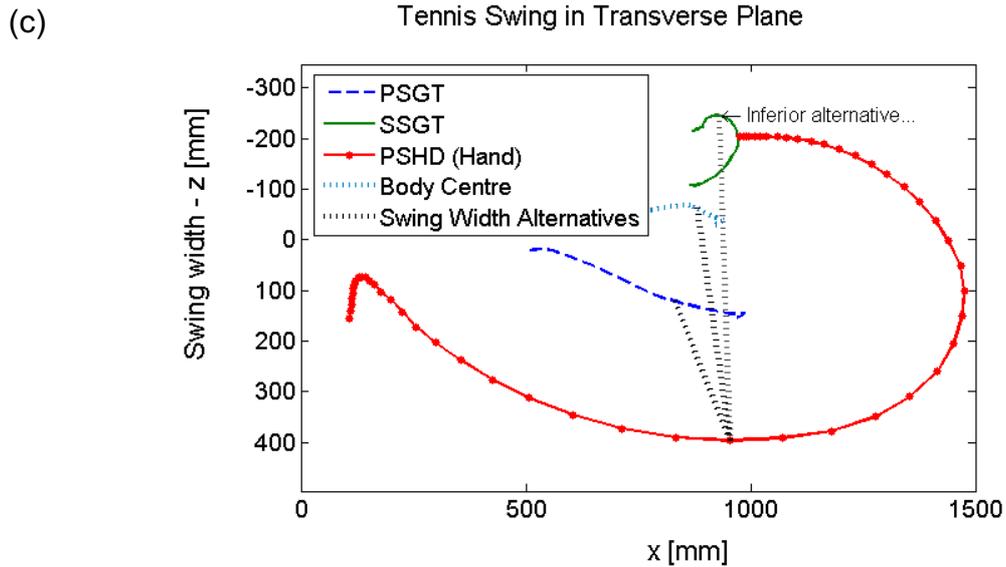

Fig. 3. Visualisation of performance (a), (b) and safety (c) parameters extraction from tennis swing data. Temporal and spatial analysis of: (a) low-to-high swing segment in sagittal view (b) finding local swing minimum as region of interest before the impact zone and (c) comparisons of algorithms alternative as knowledge discovery from machine to human.

Fig. 3 shows: (a) sagittal view of wrist low-to-high angle approximation for low sampling frequency and (b) transverse view of swing width between the wrist ('PSHD') and hip markers: great trochanter ('PSGT' and 'SSGT') and derived body/hip centre. The stance angle extraction method (Boris Bačić, 2016b) operated on estimating the average angle of the tip of the shoes relative to the intended target line. After motion data transformation for each *coaching rule* (CR) classification module, the extracted values (low-to-high angle, relative swing distance and relative stance angles) and the swing type (forehand and backhand) were linearly normalised for the classification model into a range of values between [0,…,1].

Classification results (Table 1) showed 81% accuracy compared to the human expert for 'low-to-high' CR which was due to the low sampling frequency. This made it difficult to determine the curve properties of the racquet and wrist or their lowest point relative to the body and stance. To improve classification of 'low-to-high' CR, the PTCS or exergaming capture systems operating on lower frequencies would require higher sampling frequency than stance computing. Regarding the 'swing-width' CR, playing hand–near hip relative distance algorithm was inferior compared to the algorithm variations taking into account the body (computed virtual body centre or the opposite part of the pelvis compared to playing hand that for sometimes may be occluded to the observer). Given better machine learning classification performance, this may indicate that swing width should be explained to coaches and players as more holistic assessment rather than just wrist–near hip distance based assessment.

Table 1. Leave-one-out cross-validation results using Evolving Clustering Function (ECF) for coaching rules classifier models.

| Coaching rule | Variation | Classification results | ECF epochs |
|---|---|---|---|
| Stance | Square | 91% OA; (39/43 correct) | 4 |
| | Semi-open | 91% OA; (39/43 correct) | |
| 'Low-to-high' | | 81% OA; (35/43 correct) | 2 |
| Swing width | hand – leading hip | 81% OA; (35/43 correct) | 2 |
| | hand – body centre | 91% OA; (39/43 correct) | |
| | hand – rear hip | 91% OA; (39/43 correct) | |

OA = Overall accuracy; ECF membership functions = 3, 2 and 1 did not change OA.

Given the subjective nature of stance assessment decision boundaries and the need for skill-level and coaching scenarios, the design decision was to develop two separate classification modules (Fig. 4 a) – one for 'Square Stance' (SqS) and another for 'Semi-open Stance' (SoS).

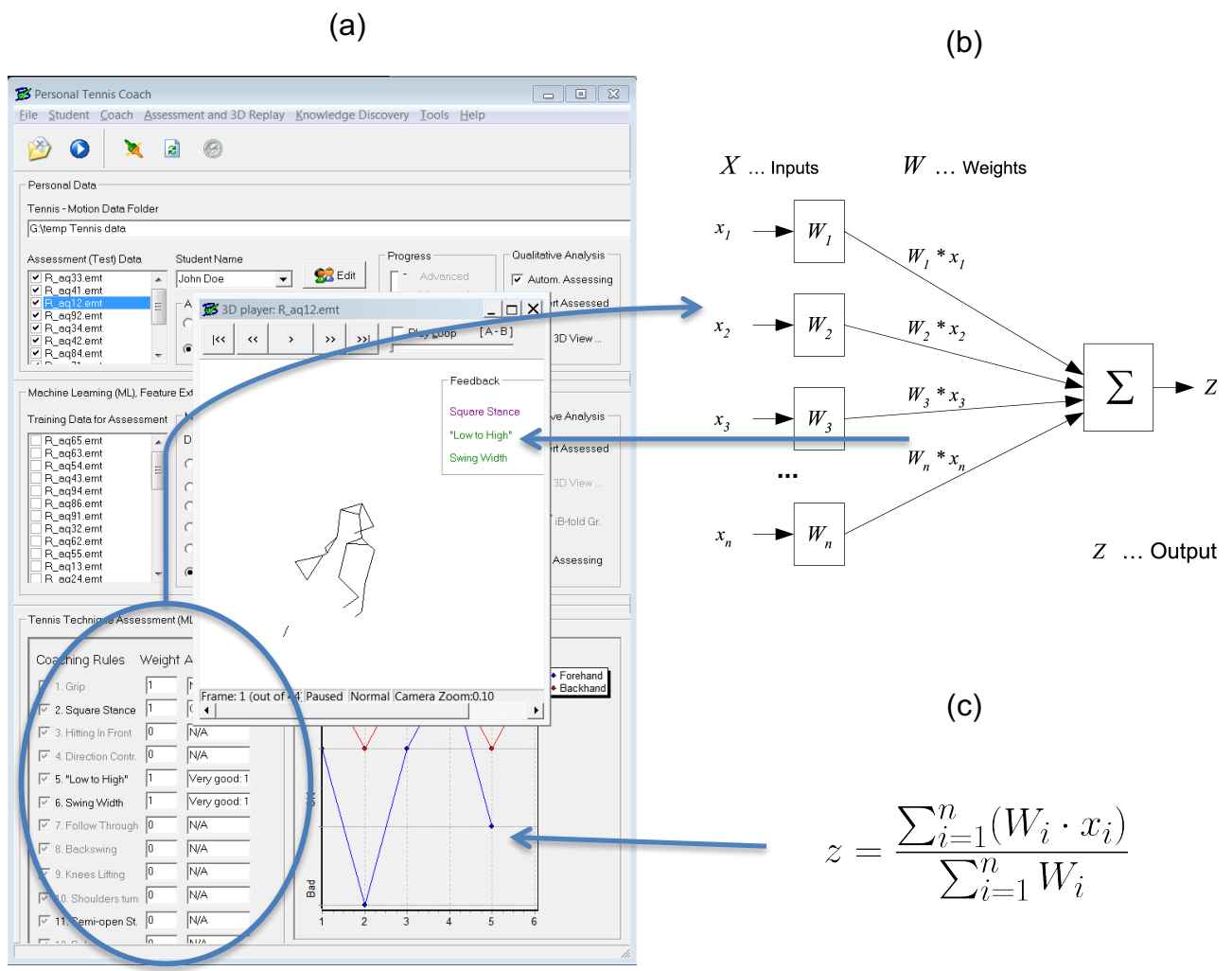

Fig. 4. User interface, and weighted orchestration of assessment modules $W_i$ to produce swing evaluation as overall assessment $Z$, and feedback as sorted list of colour-coded attention cues (from worst to best).

Adaptive weighted orchestration of assessment modules in PTCS (Fig. 4) allowed incremental development and integration of new CR modules (depicted in grey) without changing the existing software infrastructures. Fig. 4 shows the user-specified weighted assessment concept that leverages an independent and incremental modular design is implemented via weighted assessment criteria for coaching rules. Swing evaluation as weighted assessment could correspond to the CS for global, group or personal idiosyncrasies such as skill level. The formula (Fig. 4 c) also supported the orchestration concept that enabled the scalable, incremental and collective assessment operation $Z$ of CR modules that could be used for in-between-session(s) performance/technique evaluation purposes. For an end-user, the weights vector $W$ could be supplied within [0,…,1] values or assigned as a percentage or as any grading preference. The output of CR modules $x_i$ (indexed as $i = 1,…,n$) were multiplied by the weights vector $W$ matching skill level and coaching scenario for the individual player. The weighted assessment was displayed in the 3D stick figure player as colour-coded CRs that are equivalent to verbal cues and displayed immediately after the swing has been assessed. The end-user may configure to show only the worst stylistic execution aspect or multiple aspects as attention cue list (e.g. the top three sorted from worst to best).

**Discussion**

Although we were unable to generate rules that govern expert's tacit knowledge from our synthetic expert coaching system, at present, we are able to provide both holistic-global evaluation and weighted combination of assessed motion based on common-sense rules that guide qualitative assessment of human motion (Fig. 2). Furthermore, weighted combination of assessed motion also represents an adaptive mechanism for new assessment modules to be added or replaced in the future, while the coach will still be able to modify the weighted combination for given skill-level, specific drill-based training objectives or coaching scenarios. The adaptive ECF algorithm implemented in Neucom (Goh, Song, & Kasabov, 2004; Song et al., 2008) that was originally used in high-dimensional gene research worked well with relatively small tennis data set containing a balanced number of good and bad tennis groundstrokes that are typical for beginners to intermediate skill levels. The leave-one-out cross validation method used in the experiments was considered appropriate to address the scientific rigour involving modelling and analysis of relatively small data sets. Future work with large datasets will also investigate other supervised learning techniques applicable to big data such as Vapnik's *transductive* reasoning and experts' review of subset of data combined with validation scores (M M Patching et al., 2015).

Based on this study, technological innovations in near-future ACST, exergames and rehabilitation devices are also expected to utilise diverse data sources such as video, computer vision processing (e.g. utilising deep learning ANNs), various inertial sensors that are currently used for movement patterns, swing detection, pattern classification and various motion analysis (Cao, 2017; Manghisi et al., 2017; Whiteside, Cant, Connolly, & Reid, 2017). Given the general trend of wearable technology manufacturers not allowing access to raw data or integration libraries to end users but rather claiming the ownership of data, such end-user agreements should raise questions regarding ethical and legal challenges particularly in terms of privacy and intellectual ownerships (Bačić, 2018; Socolow, 2016, 2017). In the authors' view, in the future we may expect: (1) open source hardware initiative for wearable and sport technology similar to open source software licensing initiatives to promote ACST research and improve engagement from broad academic community; (2) web services, exergaming and ACST development involving both small start-up companies and international corporations such as Google's translate services to process video and raw data inputs to detect human activity and produce cues, augmented visualisations, feedback with references and recommend intervention; (3) advancement of privacy-preserving technologies (B. Bačić, Meng, & Chan, 2017; Chan & Bačić, 2018) for online coaching, healthcare, smart homes and elderly care; (4) advancements in surveillance systems based on movement signature person identification operating and accessing large data sets; and (5) coach's assessments on captured data and assessment model design to be treated as intellectual property. The next-generation of exergames and ACST should be able to: (1) capture motion data from various sources (Cao, Simon, Wei, & Sheikh, 2017; Lightman, 2016; MathWorks, 2016); (2) process complex motion pattern in on-line and off-line fashion; (3) provide automated analysis of human motion performance for given tasks and associated objectives (e.g. efficiency, efficacy, safety, consistency, error and accuracy rates); (4) provide feedback; and (5) suggest intervention for improvements. ACST application in coaching and rehabilitation include: automating coaching practices; off-loading cognitive tasks performed by a coach (e.g. common errors, safety monitoring, and progress management support); reduction of bias, disagreements, inconsistencies and fear of reinjures; and support for semi-supervised and self-coaching.

## Conclusions

The presented system demonstrated that it was possible to generate feedback consisting of elements of safety and performance to help motor learning or improving complex sport-specific technique such as tennis swing. As multi- and inter-disciplinary scientific contribution, the demonstrated proof-of-concept system was able to capture expert insights into a computer model that and reproduce qualitative diagnostics on previously unseen motion data similar to human reasoning (above 80%). Demonstrated automated diagnostic feedback was associated with: (1) subjective expert's assessment and feedback containing abstract and descriptive common-sense rules associated with performance and safety; (2) critical features of sport-specific human movement sequence that can also operate with AI-based systems; and (3) common errors and attention cues. For the safety aspect of the relative width of a tennis swing, the prototype demonstrated improved assessment (from 81% to 91%) when taking into account occluded parts of the pelvis on the same data, which as evidence is also considered as knowledge discovery from data that can inform coaching practice. To address the need for life-long adaptive learning (for sports such is tennis), the system and classifier models have properties that computational intelligence consider as adaptive, evolving, and life-long learning from initially small training data.

    We demonstrated that capability of qualitative intuitive feedback in natural language (e.g. as verbal attention cueing) and visual augmented elements (e.g. 3D replay) with assessed relative performance over time, were aligned with coaching practice. This has potential to improve the end user's technique more than using existing coaching communication of only quantitative outcomes of observed movements. Machine-generated qualitative analysis for coaching feedback of complex motion patterns to improve motor function, control and technique is commonly applicable to a range of sport disciplines and rehabilitation scenarios. The

underlying technology foundation covers existing and future motion capture devices capable of generating increasingly large data set such as: wearables, sport/action cameras, mobile phones, sensors attached to sport equipment, game and exergame controllers, EEG, EMG, functional rehabilitation devices, intelligent prosthetics and exoskeleton control design.

Combining *augmented coaching systems and technology* (ACST) with near-future advancements of exergaming and immersive reality offer new prospective for aging population, rehabilitation patients, sport participants, and those who aspire to healthier and more active lifestyle. As such, developments of autonomous augmented coaching systems and technology (ACST) represent the opportunity to strengthen the links between exercise and health. Future work will extend to: (1) incremental modelling of other coaching rules for tennis; (2) technology transfer to other areas; (3) data fusion from diverse motion data sources; (4) distributed multi-platform data processing; and (5) active monitoring devices associated with rehabilitation, coaching, ergonomics and general well-being.

**Acknowledgements**

We wish to express appreciation to Professor Stephen MacDonell for his support, and Professor Barry Wilson and Professor Will Hopkins for helpful feedback before shaping this article. We also wish to acknowledge the contribution of Professor Nik Kasabov, who directed the development and shared the Matlab code of NeuCom (www.theneucom.com) used as classifier in this study. Tennis data were obtained in collaboration with the Peharec polyclinic for physical therapy and rehabilitation, Pula (Croatia). The tennis swing dataset acquisition and execution performance was scrutinised and assessed by Mr. Petar Bačić (biomechanics lab specialist and professional tennis coach with international experience). In addition, we wish to acknowledge New Zealand tennis coaches, Kevin Woolcott and Shelley Bryce for their valuable comments regarding the user interaction aspects of the prototype and the tennis data set assessment. The study has been mainly supported by the author's discretionary transitional research funding from AUT with no external financial support.

**Disclosure Statement**

No potential conflict of interest was reported by the authors.

Corresponding Author: Dr Boris Bačić, Level 1, AUT Tower, 2-14 Wakefield St, Auckland 1010, New Zealand